\documentstyle[aps,psfig,multicol]{revtex}

\begin{document}
\draft

\begin{center}
{\bf\it Perspectives: Superconductivity}

\vspace{0.1cm}

{\bf The Race to Beat the Cuprates}

\vspace{0.1cm}

{[Science 239, 2410 (2001)]}

\vspace{0.1cm}

{\small Elbio Dagotto}

\vspace{0.1cm}

{\small\it Department of Physics and National High Magnetic Field Lab, Florida State University,
Tallahassee, FL 32306, USA.}\\
{\small\it E-mail: dagotto@magnet.fsu.edu}
\end{center}

\vspace{0.5cm}

  Superconductors are materials that lose all electrical resistance 
  below a specific temperature, known as the critical temperature ($T_c$). Large-scale
  applications, for example, in superconducting cables, require materials with high
  (ideally room temperature) $T_c$'s, but most superconductors have very low $T_c$'s, 
  typically a few kelvin or less. The discovery of a layered copper oxide (cuprate)
  with a $T_c$ of 38 K (see panel A in the first figure) in 1986 [1] raised hopes that
  high temperature superconductivity might be within reach. By 1993, cuprate $T_c$'s of
  133 K at ambient pressure had been achieved [2,3], but efforts to further increase
  cuprate $T_c$'s have not been fruitful. Two reports by Sch\"on {\it et al.} [4,5]
  in the current issue of science --applying a similar technique to two very different materials-- 
  drastically alter the perception that planar cuprates are the only route 
  to high temperature superconductivity.

  Sch\"on {\it et al.} use a field-effect device introduced in previous investigations to
  transform insulating compounds into metals [6]. On page 2430, they show that 
  copper oxide materials with a ladder structure (panel B in the first figure) can be
  superconducting [4],  even without the high pressure applied in previous studies of
  related compounds. Even more spectacularly, they report on page 2432 that the $T_c$ of
  a noncuprate molecular materials, $C_{60}$ (panel C in the first figure), 
  known before to superconduct at $52~K$ upon hole doping [7], can be raised 
  by hole doping with  intercalated CHBr$_3$ to $117~K$ [5], not far from the 
  cuprate record. Simple extrapolations suggest that the $T_c$ could be increased 
  even further, effectively ending the dominance of cuprates in the high-$T_c$ arena.

  The idea behind the studies is conceptually simple. Field-effect doping
  exploits the fact that under a strong, static electric field,
  charge (electrons or holes) will accumulate at the surface of the material, 
  effectively modifying the electronic density in that region. This is necessary 
  to stabilize superconductors away from nominally insulating compositions. The 
  dielectric portion of the field-effect device must be able to sustain electric 
  fields large enough to induce a sufficient number of holes per atom or molecule 
  for the material under study to become superconducting. In addition, the interface 
  with the studied material must be as perfect as possible. 
  Doping through a field-effect device [4,5] avoids imperfections that
  cause the system to deviate locally from its average properties. Such imperfections
  are inevitably induced by chemical doping. Disorder has not been seriously considered 
  by most cuprate high-$T_c$ theorists, but its important role is slowly emerging. Some 
  phase diagrams of cuprates may have to be redrawn when doping is introduced through 
  a field-effect device [8].

\noindent 
\begin{figure}
\begin {center}
\mbox{\psfig{figure=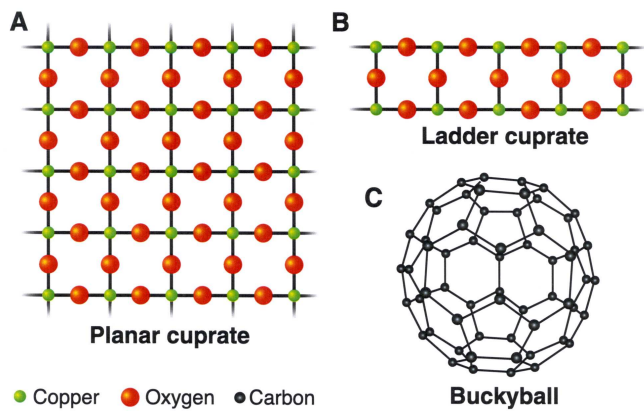,height=2.5in}}
\end {center}
\caption{
{\bf The structures of superconductors} ({\bf A}) Copper oxide plane,
{\bf (B)} copper oxide ladder,and {\bf (C)} $C_{60}$ molecule. Ladders of copper
and oxygen atoms, as shown in (B), form spontaneously in some compounds.
}
\end{figure}

  In ladder cuprates, electrons are more likely to move along the legs of the ladder,
  rendering the material quasi-one-dimensional. However, the ladder rungs are 
  also very important: They induce an effective attraction between carriers, in this 
  case holes, that leads to superconductivity [9]. 
  Superconductivity in a ladder material was first observed experimentally in 1996,
  when a $T_c$ of $13~K$ was reported [10]. However, the superconducting state was 
  stabilized with a high pressure of about 3~GPa. Ambient pressure ladder 
  superconductivity, although searched for extensively, was not observed until now. 

  The previous negative results suggested that high pressure may transform the
  ladders into anisotropic two-dimensional systems (similar to the planar cuprates) by
  reducing interatomic distances [11]. However, Sch\"on {\it et al.} show that ladders
  can become superconducting without high pressure, simply by a different doping 
  procedure than previously used. Cuprate superconductivity is thus not unique 
  to two-dimensional structures but exists in ladders 
  as well, with similar copper and oxygen building blocks but a 
  qualitatively different atomic arrangement.  

  The ladder compounds are conceptually important because they provide the
  only known superconducting copper oxide without a square lattice. The hypothesis
  that the ladder compounds are anisotropic two-dimensional systems appears
  difficult to sustain in view of the discovery reported in [4]. Furthermore,
  the resistivity at optimal doping (when $T_c$ is the highest) is linear with 
  temperature [4]. This behavior, observed also in non-superconducting ladders [11], 
  was previously believed to be a unique signature of the exotic properties of 
  high-$T_c$ planar cuprates.

  Crystalline $C_{60}$ is normally an insulator, but in 1991, it was shown
  that electron-doped fullerenes are superconducting [12]. Recently, 
  the $T_c$ in these compounds was raised to $52~K$ by field-effect hole
  doping, suggesting that $T_c$ could be raised further by increasing the 
  intermolecular distance -- a quantity that was found to be almost linearly 
  related to $T_c$ [7].
 
  The new results [5] confirm these expectations. It is widely believed that 
  hole-doped $C_{60}$ follows the standard model of superconductivity in which 
  phonons (vibrations of the lattice and molecules) provide the source of attraction 
  between carriers for pair formation and concomitant zero resistance. In fullerenes,
  high-energy intramolecular phonons are available to mediate the pairing. As 
  the distance between molecules increases, the overlap of electronic wavefunctions 
  decreases. As a result, the electronic bands narrow and the electronic
  density of states at the Fermi level ($E_F$) increases. These effects, 
  supplemented by a substantial electron-phonon coupling ($\lambda$), appear to
  determine to a large extent the high value of the $T_c$. Smaller $C_{36}$ 
  fullerenes are expected to have a larger $\lambda$
  than $C_{60}$ [13], suggesting another route to higher $T_c$'s.

  These arguments are persuasive and likely correct, but possible 
  electronic pairing mechanisms should also be considered. Electron-electron 
  interactions are characterized by an energy scale much larger than that 
  of phonons and are more likely to generate high-$T_c$ 
  behavior. Intramolecular pairing of whatever origin -- phononic or electronic -- 
  may produce the same local effective attraction, usually referred to as ``-$U$''. 
  Denoting by $t$ the amplitude for electron hopping between $C_{60}$ molecules, 
  in simple models for superconductivity the reduction of $t$ at fixed $|U|$ leads to an
  increase of $T_c$ in weak coupling [14], as in the present experiments [5],
  where the hopping is regulated by intercalating small molecules [15].
  On the other hand, if $T_c$'s of more than 100 K can be achieved in fullerenes 
  with just phonons, then the relevance of phononic mechanisms for cuprates should be 
  reconsidered. Has nature given us only one way to induce high temperature
  superconductivity, after all?

\begin{figure}
\begin {center}
\mbox{\psfig{figure=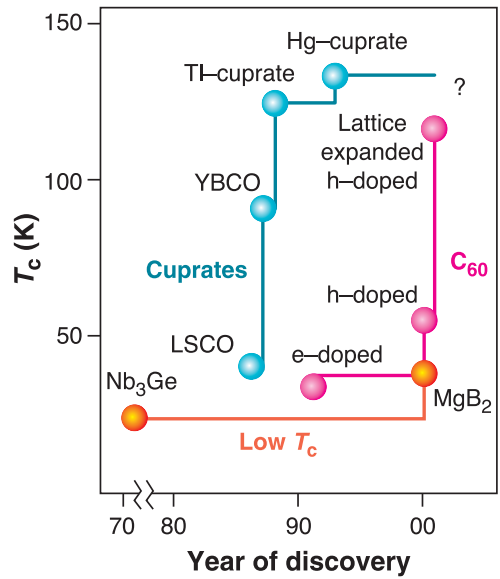,height=3.in}}
\end {center}
\caption{
{\bf Head-to-head race,} $T_c$ versus year of discovery for some 
superconducting materials. Orange, representative low-$T_c$ compounds,
which held the $T_c$ record before cuprates and fullerenes were discovered. 
Light blue, representative planar cuprates.
Magenta, representative fullerenes, with the highest $T_c$ to date reported in [4].
}
\end{figure}

  Superconductivity in field-effect doped materials is effectively
  two-dimensional, sandwiched between the undoped bulk material and the 
  dielectric oxide. Is this effective lower dimensionality 
  crucial for the high $T_c$ obtained? Results for electron-doped 
  fullerenes suggest otherwise --- bulk- and field-effect-doped compounds 
  have similar $T_c$'s [7] --- but the question is worth investigating.  
  Correlating $T_c$ with the electronic density of states at $E_F$  would 
  also allow for a more intuitive understanding of the results. The mechanism of 
  increasing the density of states by reducing the bandwidth through the expansion
  of the lattice seems simplistic but appears to work. In theoretical studies of 
  superconducting cuprates, band narrowing was caused by the antiferromagnetic
  background in which the holes are immersed. In this context, optimal doping is
  naturally associated with a peak in the density of states [16].

 Through increasing the $C_{60}$ lattice constant by 1\% [5]
 or improvements of the field-effect device, it may be possible to induce
 a $T_c$ above 133 K, the ambient pressure record for cuprates (see the second figure). 
 However, the work on ladders [4] shows that cuprates can now 
 also be doped by the field-effect device. An exciting organic vs inorganic race 
 toward room-temperature superconductivity may be about to begin. If so, then field-effect 
 doping will likely play a fundamental role.

\begin{center}
{\bf REFERENCES}
\end{center}

\begin{description}

\item{1.} J. G. Bednorz and K. A. Mueller, Z. Phys. B{\bf 64}, 189 (1986).

\item{2.} A. Schilling {\it et al.}, Nature {\bf 363}, 56 (1993). 

\item{3.} M. Nunez-Regueiro {\it et al.}, Science {\bf 262}, 97 (1993).

\item{4.} J. H. Sch\"on {\it et al.}, Science {\bf 293}, 2430 (2001).

\item{5.} J. H. Sch\"on, Ch. Kloc, and B. Batlogg, Science {\bf 293}, 2432 (2001).

\item{6.} J. H. Sch\"on {\it et al.}, Science {\bf 288}, 656 (2000).

\item{7.} J. H. Sch\"on {\it et al.}, Nature {\bf 408}, 549 (2000).

\item{8.} J. Burgy {\it et al.}, cond-mat/0107300, to appear in Phys. Rev. Lett., 
and references therein.

\item{9.} E. Dagotto, J. Riera and D. Scalapino, Phys. Rev. B{\bf 45}, 5744 (1992).

\item{10.} M. Uehara {\it et al.}, J. Phys. Soc. Jpn.{\bf 65}, 2764 (1996).

\item{11.} T. Nagata {\it et al.}, Phys. Rev. Lett. {\bf 81}, 1090 (1998).

\item{12.} A. F. Hebbard {\it et al.}, Nature {\bf 350}, 600 (1991).

\item{13.} M. C\^ot\'e {\it et al.}, Phys. Rev. Lett. {\bf 81}, 697 (1998).

\item{14.} R. Scalettar {\it et al.}, Phys. Rev. Lett. {\bf 62}, 1407 (1989).

\item{15.} The statement that $T_c$ may be increased by increasing 
 the lattice spacing, since the electronic pair binding is an 
 intramolecular property, already appeared in S. Chakravarty, M. 
 Gelfand, and S. Kivelson, Science {\bf 254}, 970 (1991). See also
 P. Lammert {\it et al.}, Phys. Rev. Lett. {\bf 74}, 996 (1995)
 and references therein. The author apologizes for this omission in 
 the present Science article.

\item{16.} A. Nazarenko {\it et al.}, Phys. Rev. Lett. {\bf 74}, 310 (1995).

\end{description}
\end{document}